\documentclass[aps,showpacs,twocolumn,superscriptaddress,nofootinbib,tightenlines,prl]{revtex4}
\usepackage{graphicx}
\usepackage{amsmath}
\usepackage{bbm}

\begin{document}

\title{Momentum vortices on pairs production by two counter-rotating fields}

\author{Z. L. Li}
\affiliation{School of Science, China University of Mining and Technology, Beijing 100083, China}
\author{Y. J. Li}
\affiliation{School of Science, China University of Mining and Technology, Beijing 100083, China}
\author{B. S. Xie\footnote{Corresponding author. bsxie@bnu.edu.cn}}
\affiliation{Key Laboratory of Beam Technology and Materials Modification of the Ministry of Education, College of Nuclear Science and Technology, Beijing Normal University, Beijing 100875, China}
\affiliation{Beijing Radiation Center, Beijing 100875, China}

\date{\today}

\begin{abstract}
Multiphoton pair production is investigated by focusing on the momentum structures of produced pairs in the polarization plane for the two circularly polarized fields. Upon the momentum spectra, different from the concentric rings with the familiar Ramsey interference fringes for the same handedness, however, the obvious vortex structures are found constituted by the Archimedean spirals for two opposite handedness fields. The underlying physical reasons are analyzed and discussed. It is also found that the vortex patterns are sensitive to the relative carrier envelope phase, the time delay, and the handedness of two fields, which can be used to detect the applied laser field characteristics as a probe way.
\end{abstract}

\maketitle

\textit{Introduction.}-An unstable vacuum in the presence of strong fields can decay into electron-positron pairs \cite{Sauter,Heisenberg,Schwinger,Piazza1}. For pair production two different mechanisms have been identified as the nonperturbative Schwinger mechanism corresponding to $\gamma\ll1$ and the perturbative multiphoton pair production having $\gamma\gg1$, respectively. Here $\gamma=m\omega/eE_0$ is the well-known Keldysh adiabaticity parameter \cite{Keldysh}, where $m$ and $-e$ is the electron mass and charge, $\omega$ and $E_0$ are the frequency and strength of external electric fields, respectively. The multiphoton pair production resulting from the nonlinear Breit-Wheeler process \cite{Reiss1962,Nikishov1964} after a nonlinear Compton scattering \cite{Liang2014} has been accomplished experimentally more than a decade ago at the Stanford Linear Accelerator Center via the collisions of a 46.6 GeV electron beam with an intense optical laser pulse \cite{Burke}. However, due to the participation of electron, the above multiphoton pair production does not occur only in a laser field. Therefore, multiphoton pair production from a vacuum in a pure laser field is being explored \cite{Abdukerim2013,Kohlfurst2014,Blinne2014,ZlliEPL2015,ZlliPRD2015,Fillion-Gourdeau2017}.

It is well known that there exist an intimate relation between the pair production and the study on atomic ionization. Recently, it has been found that the photoelectron momentum distributions in the polarization plane present vortex patterns in the multiphoton ionization with two counter-rotating fields \cite{Djiokap2015,Pengel2017}. This interesting finding inspires one to thought whether
there exist similar phenomena, i. e., do the vortices also exist in the momentum spectra of created electron-positron pairs in multiphoton regime for two counter-rotating circularly polarized electric fields? For the vortices, we know that they are quite general phenomena in nature, especially in fluid, and are widely studied in different physics areas, such as type-II superconductors \cite{Blatter1994}, atomic condensates \cite{Madison2000}, atomic and molecular ionization \cite{Macek2009}, nonlinear optics \cite{Harris1994}, plasmas \cite{Uby1995,Shukla2009} and so on. On the other hand the study associated to vortices has attracted more and more attention of the researcher since the formation of vortices often implies many deeper physical mechanisms behind the studied problems \cite{Blatter1994,Shukla2009,Salman2013,Shao2017}.

In this letter, by numerically solving the real-time Dirac-Heisenberg-Wigner formalism \cite{Bialynicki,Hebenstreit2010}, we study the momentum signatures in multiphoton pair production for two circularly polarized electric fields with a time delay, and find that indeed there exist also the vortex structures in the momentum spectra of created electron-positron pairs for two counter-rotating fields. We further explore the formation reason of vortex patterns, which is significantly different from that in atomic ionization, and uncover more important signatures of pair production in circularly polarized fields. Especially the obtained information of momenta spectra can be used to detect the fields characteristics as a probe way.

\textit{Theoretical formalism.}-The external field we focused on is a spatially homogeneous and time-dependent electric field which is composed of two circularly polarized electric fields as
\begin{eqnarray}\label{eq1}
\hspace{-0.5em}\mathbf{E}(t)& \hspace{-5em}=\hspace{-0.2em}E_1\exp\Big(-\frac{t^2}{2\tau^2}\Big)\left[
                                             \begin{array}{c}
                                               \cos(\omega t+\phi_1) \\
                                               \delta_1\sin(\omega t+\phi_1) \\
                                               0 \\
                                             \end{array}
                                           \right] \nonumber \\
&\hspace{-0.5em}+E_2\exp\Big(-\frac{(t-T)^2}{2\tau^2}\Big)\left[
                                             \begin{array}{c}
                                               \cos\big(\omega (t-T)+\phi_2\big) \\
                                               \delta_2\sin\big(\omega (t-T)+\phi_2\big) \\
                                               0 \\
                                             \end{array}
                                           \right],
\end{eqnarray}
where $E_{1,2}=E_0/\sqrt{1+\delta_{1,2}^2}$ are the amplitudes of the electric field, $\delta_{1,2}=\pm1$ are the circular polarizations, $\tau$ denotes the pulse duration, $\omega$ represent the field frequency, $\phi_{1,2}$ are the carrier envelope phases (CEPs), and $T$ is the time delay between these two circularly polarized fields. For the convenience of calculation, we set $E_0=0.1\sqrt{2}E_\mathrm{cr}$, $\omega=0.6m$, $\tau=10/m$, and $\phi_1=0$ throughout this paper unless otherwise stated. Here $E_{\mathrm{cr}}=m^2/e\sim1.32\times10^{16}\mathrm{V/cm}$ is Schwinger critical field strength, $e$ is the magnitude of electron charge (the units $\hbar=c=1$ are used).

\begin{figure}[htbp]\suppressfloats
\vskip -4cm
\includegraphics[width=8cm]{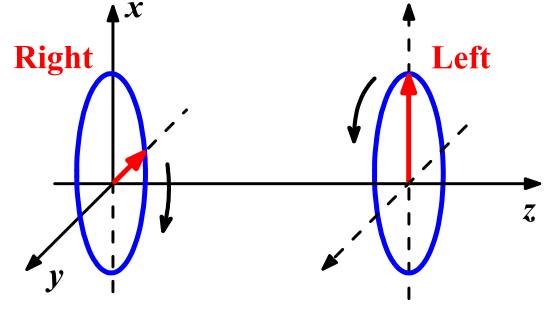}
\caption{\label{Fig1}(color online) Sketch of the electric field (\ref{eq1}). The first pulse (right ellipse) is a left-handed circularly polarized electric field ($\delta_1=1$) and the second one (left ellipse) is a right-handed circularly polarized field ($\delta_2=-1$) delayed in time by $T$. $(x,y)$ plane is the polarization plane.}
\end{figure}

We use the same method as in Ref. \cite{Blinne2016}, then the one-particle distribution function $f(\mathbf{q},t)$ can be obtained by solving
\begin{eqnarray}\label{EOM1}
\dot{f}&=&\frac{e\mathbf{E}\cdot\mathbf{v}}{2\Omega}, \nonumber\\
\dot{\mathbf{v}}&=&\frac{2}{\Omega^3}[(e\mathbf{E}\cdot\mathbf{p})\mathbf{p}
-e\mathbf{E}\Omega^2](f-1) \nonumber \\
&&-\frac{(e\mathbf{E}\cdot\mathbf{v})\mathbf{p}}{\Omega^2}
-2\mathbf{p}\times\mathbbm{a}-2m\mathbbm{t}, \\
\dot{\mathbbm{a}}&=&-2\mathbf{p}\times\mathbf{v}, \nonumber\\
\dot{\mathbbm{t}}&=&\frac{2}{m}[m^2\mathbf{v}
-(\mathbf{p}\cdot\mathbf{v})\mathbf{p}], \nonumber
\end{eqnarray}
with the initial conditions: $f(\mathbf{q},-\infty)=0, \mathbf{v}(\mathbf{q},-\infty)=\mathbbm{a}(\mathbf{q},-\infty)
=\mathbbm{t}(\mathbf{q},-\infty)=\mathbf{0}$. Here $\mathbf{v}(\mathbf{q},t)$ is an auxiliary quantity, $\mathbbm{a}(\mathbf{q},t)$ and $\mathbbm{t}(\mathbf{q},t)$ are the Wigner components, $\Omega(\mathbf{q},t)=\sqrt{m^2+[\mathbf{q}-e\mathbf{A}(t)]^2}$ is the total energy of electrons, $\mathbf{q}$ denotes the canonical momentum,  $\mathbf{A}(t)$ is the vector potential of the electric field $\mathbf{E}(t)$, $\mathbf{p}(t)$ is the kinetic momentum and has $\mathbf{p}(t)=\mathbf{q}-e\mathbf{A}(t)$. Moreover, we can also get the number density of created pairs $n(+\infty)$ by integrating the distribution function over full momenta at $t\rightarrow+\infty$, i.e.,
\begin{equation}\label{numberdensity}
n(+\infty)=\int\frac{d^3q}{(2\pi)^3}f(\mathbf{q},+\infty).
\end{equation}

\begin{figure}[htbp]\suppressfloats
\vskip -4cm
\includegraphics[width=8cm]{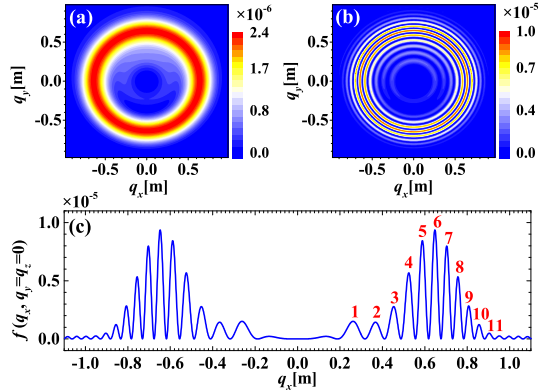}
\caption{\label{Fig2}(color online) Momentum spectra of created electron-positron pairs for different electric fields. (a) for a single left-handed circularly polarized pulse, i.e., $\delta_1=1$ and $E_2=0$. (b) and (c) for two left-handed circularly polarized fields with the time delay $T=100/m$, and $E_2=E_1=0.1E_{\mathrm{cr}}$, $\delta_2=\delta_1=1$, $\phi_2=\phi_1=0$. (a) and (b) are the momentum spectra in the polarization plane for $q_z=0$. (c) is the momentum distribution for $q_y=q_z=0$. }
\end{figure}

\textit{Numerical results and analysis.}-By solving Eqs. (\ref{EOM1}), we first calculate the momentum spectra of created electron-positron pairs for a single left-handed circularly polarized electric field and two left-handed circularly polarized (LLCP) ones, see Fig. \ref{Fig2}. From Fig. \ref{Fig2}(a), one can see that there is an obvious ring structure with a weak interference effect along negative $q_y$ at $q_x=0$ in the momentum spectrum. This pattern has been pointed out in Ref. \cite{Blinne2014} and further studied in Ref. \cite{ZlliPRD2015}. It is known that the ring is caused by $4$-photon pair production and its radius can be determined by the energy conservation equation after considering the effective mass. Furthermore, the weak interference effect can be understood from the interference between different turning points $t_p$ in the complex time plane which are obtained by solving $\Omega(\mathbf{q},t_p)=0$ for a certain $\mathbf{q}$. In fact, this interference patten in Fig. \ref{Fig2}(a) is caused by the interference between the dominant tuning points ($t_p$ closest to the real time axis) and the other ones (mainly the subdominant ones), see Fig. 7 in Ref. \cite{Blinne2016}. As the number of cycles within the pulse duration increases, more turning points will become closer to the real time axis and reach the same distance to the real time axis as the dominant ones. Consequently, the interference between these turning points makes the interference pattern in momentum spectrum become more obvious. This can be seen from the third pattern of Fig. 2 in Ref. \cite{ZlliPRD2015}. For the LLCP electric field with the time delay $T$, the momentum spectrum of created pairs is shown in Fig. \ref{Fig2}(b). It is found that there are many concentric rings and the interference effect is very obvious. Actually, this pattern is the Ramsey interference fringes which are formed by the interference between two pairs of dominant turning points near $t=0$ and $t=T$ corresponding to each pulse of the two left-handed fields. This interference effect is a bit different from that for a single left-handed circularly polarized pulse, because it is caused by the dominant turning points from two pulses rather than one. This result can be explained semiquantitatively by the Wentzel-Kramers-Brillouin (WKB) like approximation employed in Refs. \cite{Strobel2015,Blinne2016} as follows.

Because electron-positron pairs are mainly produced at the maximum of electric field, i.e., at $t=0$ and $t=T$ for the LLCP electric field, so the two pairs of turning points near $t=0$ and $t=T$ dominate pair creation process. This is similar to the case studied in Refs. \cite{Akkermans2012,ZLLiPRD2014} for two alternating-sign pulses. Each pair of turning points $t_p^{\pm}$ has a contribution of $\exp[-2\vartheta_s(\mathbf{q},t_p^+)]$ to pair production, where $\vartheta_s(\mathbf{q},t_p^+)=-\mathrm{Im}[K_s(\mathbf{q},t_p^+)]$, $K_s(\mathbf{q},t)=K_0(\mathbf{q},t)-sK_{xy}(\mathbf{q},t)$, $K_0(\mathbf{q},t)=2\int_{-\infty}^t\Omega(\mathbf{q},t')dt'$,
$K_{xy}(\mathbf{q},t)= \epsilon_\perp\int_{-\infty}^t\frac{\dot{p}_x(t')p_y(t')-\dot{p}_y(t')p_x(t')}
{\Omega(\mathbf{q},t')[p_x^2(t')+p_y^2(t')]}dt'$, $s=\pm1$ denotes the electron spin, $s=0$ represents the scalar particle, and $\epsilon_\perp=\sqrt{q_z^2+m^2}$. It is known that for a certain $\mathbf{q}$ the amplitude of pair production for the first circularly polarized electric field is $A_1=\exp[-iK_s(\mathbf{q},t_0^+)]$ and the amplitude for the second one is $A_2=\exp[-iK_s(\mathbf{q},t_T^+)]$, where $t_0$ and $t_T$ denote the turning points near $t=0$ and $t=T$. Therefore, we can get the momentum distribution function
\begin{eqnarray}\label{MD1}
f(\mathbf{q})&=&\sum_{s=\pm}|A_1+A_2|^2 \nonumber \\
&=&\sum_{s=\pm}\Big|e^{-iK_s(\mathbf{q},t_0^+)}+e^{-iK_s(\mathbf{q},t_T^+)}\Big|^2.
\end{eqnarray}
Since the two circularly polarized electric fields have the same profile, the distance of the turning points $t_T$ to the real time axis is the same as that of turning points $t_0$. Thus, for a large time delay $T$, the amplitude of pair creation for the second pulse becomes $A_2=\exp[i\theta_s(\mathbf{q})]A_1$, where $\theta_s(\mathbf{q})$ is a phase accumulated between the two pulses and $\theta_s(\mathbf{q})=\mathrm{Re}[K_s(\mathbf{q},t_T^+)
-K_s(\mathbf{q},t_0^+)]$. Then Eq. (\ref{MD1}) becomes
\begin{eqnarray}\label{MD2}
f(\mathbf{q})&=&\sum_{s=\pm}|A_1+e^{i\theta_s(\mathbf{q})}A_1|^2 \nonumber \\
&=&\sum_{s=\pm}2(1+\cos[\theta_s(\mathbf{q})])e^{-2\vartheta_s(\mathbf{q},t_0^+)} \nonumber \\
&\propto&\big\{1+\cos[\theta_0(\mathbf{q})]\big\}e^{-2\vartheta_0(\mathbf{q},t_0^+)}.
\end{eqnarray}
In general, the exact analytical solution of turning points for the electric field (\ref{eq1}) can not be obtained, for example the field we considered here. But we can still explain the interference effect in Fig. \ref{Fig2}(b) under a reasonable approximation. As the time delay $T>>\tau$, the electric field and vector potential are very small between $t=0$ and $t=T$, so we can get $\theta_0(\mathbf{q})\approx 2\sqrt{\mathbf{q}^2+m^2}T$. Then from Eq. (\ref{MD2}), one can determine the positions of the Ramsey fringes in Fig. \ref{Fig2}(b) as $|\mathbf{q}|=\sqrt{(2k\pi/2T)^2-m^2}$, $k$ is an integer. For simplicity, we consider the one-dimensional case by choosing $q_y=q_z=0$, so we have
\begin{equation}\label{qeva}
q_x^{\mathrm{eva}}=\sqrt{\Big(\frac{k\pi}{T}\Big)^2-m^2}.
\end{equation}
The one-dimensional momentum spectrum is shown in Fig. \ref{Fig2}(c) and the positions of the $11$ peaks are shown in Tab. \ref{tab1}. From Tab. \ref{tab1}, one can see that the evaluate result is in good agreement with the numerical one. The small difference between them is mainly caused by the electron spin and the electric field profile (\ref{eq1}).

\begin{table*}[!ht]\suppressfloats
  \centering
  \caption{Comparison of $q_x$ and $q_{\mathrm{eva}}$ corresponding to the $11$ peaks in Fig. \ref{Fig2}(c). $q_x$ is the numerical result, and $q_x^{\mathrm{eva}}$ is the evaluate result of Eq. (\ref{qeva}) for $k=33,34,...,43$.} \label{tab1}
\begin{tabular}{|c|c|c|c|c|c|c|c|c|c|c|c|c|c|c|c|c|}
  \hline
  $i$ & 1 & 2 & 3 & 4 & 5 & 6 & 7 & 8 & 9 & 10 & 11   \\ \hline
  $q_x$ & 0.26157 & 0.36759 & 0.45361 & 0.52463 & 0.58865 & 0.64766 & 0.70368 & 0.75569 & 0.80670 & 0.85571 & 0.90373   \\ \hline
  $q_x^{\mathrm{eva}}$ & 0.27350 & 0.37540 & 0.45719 & 0.52830 & 0.59258 & 0.65205 & 0.70793 & 0.76101 & 0.81184 & 0.86081 & 0.90823   \\ \hline
  $q_x^{\mathrm{eva}}-q_x$ & 0.01193 & 0.00781 & 0.00358 & 0.00370 & 0.00393 & 0.00439 & 0.00425 & 0.00532 & 0.00514 & 0.00510 & 0.00450   \\
  \hline
\end{tabular}
\end{table*}

\begin{figure}[htbp]\suppressfloats
\vskip -4cm
\includegraphics[width=8cm]{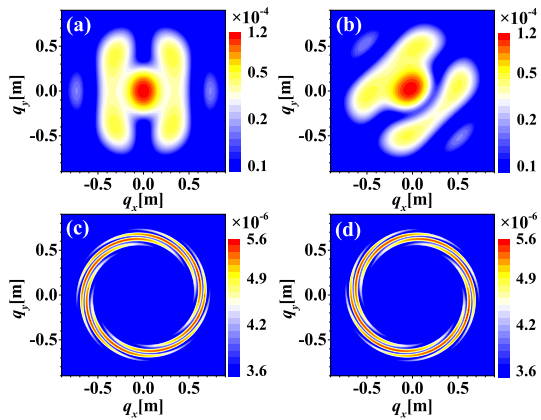}
\caption{\label{Fig3}(color online) Momentum spectra of created electron-positron pairs in the ($q_x$, $q_y$) plane (where $q_z=0$). (a) for a LRCP electric field ($\delta_1=1$ and $\delta_2=-1$) with $T=0$ and $\phi_2=0$. (b) for a LRCP electric field with $T=0$ and $\phi_2=\pi/2$; (c) for a LRCP electric field with $T=100/m$ and $\phi_2=\pi/2$. (d) for a RLCP electric field ($\delta_1=-1$ and $\delta_2=1$) with $T=100/m$ and $\phi_2=\pi/2$. Other electric field parameters are $E_2=E_1=0.1E_{\mathrm{cr}}$.}
\end{figure}

Here we consider the momentum spectra of created electron-positron pairs for the electric field composed of a left-handed (right-handed) circularly polarized electric field and a right-handed (left-handed) circularly polarized one with a time delay $T$. For convenience, this field is abbreviated as LRCP (RLCP) electric field. Figure \ref{Fig3}(a) and (b) shows the momentum spectra of created particles in the polarization plane for a LRCP electric field with $T=0$. From Eq. (\ref{eq1}), one can find that when the time delay $T=0$, the LRCP electric field is simplified to a linearly polarized electric field. So it can be seen that there is no ring structure in the momentum spectra. Moreover, by studying the effect of the relative CEP $\Delta\phi=\phi_2-\phi_1$ on the momentum distribution of created particles, it is found that the momentum distribution in the polarization plane rotates $\Delta\phi/2=\pi/4$ clockwise. This is because the polarized axis has a clockwise rotation $\pi/4$ for the LRCP field with $\phi_2=\pi/2$ compared to the one with $\phi_2=0$. Further study indicates that the rotation angle of momentum distribution created in the electric field (\ref{eq1}) is not only related to the relative CEP but also associated with the handedness of the second circularly polarized field, namely, $\delta_2\Delta\phi/2$. From this, one can see that the momentum distribution will rotate $|\Delta\phi/2|$ clockwise for $\delta_2\Delta\phi/2<0$ and counterclockwise for $\delta_2\Delta\phi/2>0$. Furthermore, we also find that the relative CEP generally rotates the momentum distribution but has little effect on the number density of created particles. These interesting results still need a further research in the future.

The momentum spectra of created pairs for a LRCP electric field with $T=100/m$ and a RLCP one are shown in Fig. \ref{Fig3}(c) and (d), respectively. Both figures exhibit a clearly visible vortex structure with approximate $c_8$ symmetry. We call the vortex whose interference fringes rotate counterclockwise (clockwise) from inside to outside as counterclockwise (clockwise) vortex. Therefore, (c) shows a counterclockwise vortex and (d) is a clockwise vortex. By the study of pair production in the LLCP electric field, we have known that the amplitude of pair production for the first electric field is $A_1=\exp[-iK_s(\mathbf{q},t_0^+)]$ and the amplitude for the second one is  $A_2=\exp[-iK_s(\mathbf{q},t_T^+)]$. To explain the formation of vortex structures in momentum spectra, however, it is convenient to express the amplitude of pair production in the spherical coordinates ($q,\theta,\varphi$), because the amplitude in spherical coordinates may give a phase factor that is able to reveal well the rotation characteristics of a circularly polarized field. Actually, the amplitude $A_1$ in spherical coordinates can be assumed as $A_1\approx\exp(i\ell\delta_1\varphi)A_0(q,\theta,\varphi)$, where $\varphi$ is the azimuthal angle and $\ell$ denotes the number of photons absorbed in multiphoton pair production. Based on the consideration of pair production in the LLCP electric field, similarly, the amplitude of pair production for the second circularly polarized field becomes $A_2\approx\exp(i\ell\delta_2\varphi)
\exp[i\theta_0(q,\theta,\varphi)]A_0(q,\theta,\varphi)$. Finally, according to Eqs. (\ref{MD1}) and (\ref{MD2}), the momentum distribution in the polarization plane is
\begin{equation}\label{MD3}
f(q,\varphi)\propto\big\{1+\cos[\theta_0(q,\varphi)+(\delta_2-\delta_1)\ell\varphi]\big\}|A_0(q,\varphi)|^2,
\end{equation}
where the polar angle $\theta$ is fixed as $\pi/2$, i.e., $q_z=0$, and $\theta_0(q,\varphi)\approx2\sqrt{q^2+m^2}T$ for $T\gg\tau$. From Eq. (\ref{MD3}), we can determine the interference fringes by
\begin{equation}\label{qmax}
q_{k'}^\mathrm{max}(\varphi)=\sqrt{\Big[\frac{2k'\pi
-(\delta_2-\delta_1)\ell\varphi}{2T}\Big]^2-m^2},
\end{equation}
where $k'$ is an integer and ensures $q_{k'}^\mathrm{max}(\varphi)$ is a real number. One can see that the above equation defines Archimedean spirals with $2\ell$-start spiral in the polarization plane. For a LRCP electric field ($\delta_1=-\delta_2=1$), equation (\ref{qmax}) shows a counterclockwise vortex by choosing $k'<0$, while for a RLCP electric field ($\delta_1=-\delta_2=-1$), it gives a clockwise vortex by choosing $k'>0$. Since Fig. \ref{Fig3}(c) and (d) are the momentum spectra for the pair production by absorbing $4$ photons, each of them shows an $eight$-start spiral vortex pattern. Moreover, it can also be found that for a LLCP or RRCP electric field ($\delta_1=\delta_2=\pm1$), the factor in the curly braces of Eq. (\ref{MD3}) is reduced to the one in the curly braces of Eq. (\ref{MD2}) and there are no vortex patterns in the momentum spectra. These results indicate that the formation of vortex structures is caused by the difference of quantum phase factors between the left-handed circularly polarized electric field and the right-handed one. In fact, the occurrence of phase factor in the amplitude of pair production can be understood physically. It is known that a circularly polarized field has a spin angular momentum $\pm1$, so each photon will carry a spin angular momentum $\pm1$ \cite{Beth1936}. Based on the conservation of angular momentum, the absorbtion of each photon can contribute a phase factor $\exp(i\delta_0\phi)$ to the amplitude of pair production. Here $\delta_0=\pm1$ denotes the polarization value of a circularly polarized field. Therefore, the phase factor in the amplitude of pair production becomes $\exp(i\ell\delta_0\varphi)$ for $\ell$-photon pair production. Figure \ref{Fig3}(c) and (d) correspond to $4$-photon pair production, so we have $\ell=4$ and the phase factor is $\exp(4i\delta_0\varphi)$ for each circularly polarized field. This causes an $eight$-start spiral vortex patterns in the momentum spectra. Correspondingly, according to the spirals number of the vortices, one can also determine the number of photons absorbed in pair production.

\begin{figure}[htbp]\suppressfloats
\vskip -4cm
\includegraphics[width=8cm]{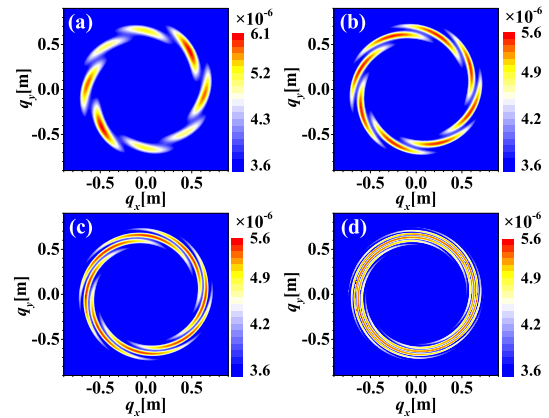}
\caption{\label{Fig4}(color online) Momentum spectra of created particles in the polarization plane (where $q_z=0$) for a LRCP electric field with different time delays. (a) for $T=30/m$, (b) for $T=50/m$, (c) for $T=80/m$, and (d) for $T=150/m$. Other electric field parameters are $E_2=E_1=0.1E_{\mathrm{cr}}$ and $\phi_2=\pi/2$.}
\end{figure}

In addition, we also investigate the effects of the time delay $T$ on the vortex structures in the momentum spectra and show the results in Fig. \ref{Fig4}. One can clearly see that the Archimedean spirals become longer and more slender with the absolute value of $T$ increases. This phenomenon can be explained by Eq. (\ref{qmax}). We rewrite this equation as $\varphi(q)=\big(2k'\pi-2\sqrt{q^2+m^2}T\big)/(\delta_2-\delta_1)\ell$. Then the absolute value of the derivative of the above equation with respect to $q$ is $|\mathrm{d}\varphi(q)/\mathrm{d}q|=|2T/(\delta_2-\delta_1)\ell|\cdot (q/\sqrt{q^2+m^2})$. This indicates that the increase of $\varphi$ with $q$ becomes more rapid for a large $|T|$ than a small one. It is also found that the vortices shown in Fig. \ref{Fig4} are distributed approximately in the same scope of $q$, i.e.,  $q_{\mathrm{in}}\sim q_{\mathrm{out}}$, because the momentum $q$ is also constrained by the energy conservation equation $2\sqrt{q^2+m^2}\approx\ell\omega$. So the corresponding increment of $\varphi$ for a large $|T|$ is greater than that for a small one. Finally, one can find that the length of the spirals increases with the time delay. Furthermore, the elongation of spirals makes vortex structures denser and the width of spirals become narrow. By calculating the number density of created electron-positron pairs in the polarization plane with Eq. (\ref{numberdensity}), we also explore the effect of the time delay on the particle number density, and find that when the two circularly polarized electric fields with different handedness are well separate, for example, $T\geq5\tau$, the particle number density will remain constant as the time delay increases. This result suggests that for each spiral of the vortex, the time delay can increase its length and reduce its width but does not change its particle number density.

\textit{Conclusions and discusses.}-In summary, in this letter, we investigate the pair production in a time-varying electric field composed of two circularly polarized electric fields with a time delay. It is found that for a LLCP or RRCP electric field with nonzero time delay, the momentum spectra exhibit concentric rings, while for a LRCP or RLCP electric field, the momentum spectra show obvious vortex patterns. By means of the WKB like approximation, we analyze these results semiquantitatively. The analysis show that the ring structures are the Ramsey interference fringes which are caused by the interference effect between two turning points corresponding to each circularly polarized field. And the vortex patterns are mainly caused by the different handedness of these two circularly polarized fields. Physically, the formation reason of vortex structures in the momentum spectra is that the photons absorbed in multiphoton pair production carry different spin angular momentum for circularly polarized fields with different handedness. We also study the effects of the relative CEP, the time delay, and the handedness on the vortices in the momentum spectra. It shows that the relative CEP $\Delta\phi$ can rotate the momentum distribution $|\Delta\phi/2|$ and its rotation direction depends on the polarization value of the second circularly polarized field. However, it has little effect on the number density of created particles. The time delay in a LRCP or RLCP electric field can change the length and width of vortex spirals, but it will not affect the particle number density when the two circularly polarized electric fields are well separate. The different handedness of these two circularly polarized fields determine the rotation direction of the vortices. By the way, it also finds that the radial width of vortex patterns, $\Delta q=q_{\mathrm{out}}-q_{\mathrm{in}}$, decreases with the pulse duration of laser field.

Based on these results, it is possible to use the information of created momentum spectra as a probe way to determine the parameters of the fields such as the CEP, the handedness and the pulse duration. In particular, according to Eq. (\ref{qmax}), the time delay between two electric fields can be evaluated by analyzing the vortex structures. Certainly beside that the present study deepen the understanding of pair production in circularly polarized fields, moreover, some questions worthy of further investigation are opened. For example, the complete quantitative interpretation of the formation of vortex structures in momentum spectra, the deep understanding of the effect of the relative CEP between two fields on pair production, the effect of two counter-rotating elliptic polarization fields, the possibility of the existence of $odd$-start spiral vortex patterns and so on.

\textit{Acknowledgements.} This work was supported by the National Natural Science Foundation of China (NSFC) under Grant Nos. 11475026 and the Fundamental Research Funds for the Central Universities.

\end{document}